\documentclass[preprint,onecolumn,amsmath,aps]{revtex4}

\usepackage{graphicx}
\usepackage[dvips]{color}
{}

\newcommand{\BAN}{\ensuremath{B_{1g}\,}}
\newcommand{\BN}{\ensuremath{B_{2g}\,}}
\newcommand{\OmAN}{\ensuremath{\Omega_{B_{1g}}\,}}
\newcommand{\OmN}{\ensuremath{\Omega_{B_{2g}}\,}}
\newcommand\Deltam{\ensuremath{\Delta_{max}\,}}
\newcommand{\ZAN}{\ensuremath{Z\Lambda_{AN}\,}}
\newcommand{\ZN}{\ensuremath{Z\Lambda_{N}\,}}
\newcommand{\vD}{\ensuremath{v_\Delta\,}}
\newcommand{\phiN}{\ensuremath{\phi_{N}\,}}
\newcommand{\cxcy}{\ensuremath{\cos k_x-\cos k_y\,}}
\newcommand{\arc}{\ensuremath{f_c\,}}

\begin{document}

\title{Suppressed antinodal coherence with a single
d-wave superconducting gap leads to two energy scales\\ in underdoped cuprates}

\author{S. Blanc, Y. Gallais, M. Cazayous, M. A. Measson and A. Sacuto}
\affiliation{Laboratoire Mat\'eriaux et Ph\'enom$\grave{e}$nes Quantiques (UMR 7162 CNRS),
Universit\'e Paris Diderot-Paris 7, Bat. Condorcet, 75205 Paris Cedex 13, France}

\author{A. Georges}
\affiliation{Centre de Physique Th{\'e}orique, CNRS, Ecole
Polytechnique, 91128 Palaiseau Cedex, France}
\affiliation{Coll\`ege de France, 11 place Marcelin Berthelot, 75005 Paris, France}

\author{G.D. Gu, J.S. Wen and Z.J. Xu}
\affiliation {Matter Physics and Materials Science, Brookhaven National Laboratory (BNL), Upton, NY 11973, USA}.

\author{D. Colson}
\affiliation{Service de Physique de l'Etat Condens\'{e}, CEA-Saclay, 91191 Gif-sur-Yvette, France}

\date{\today}
\maketitle
{\bf
Conventional superconductors are characterized by a single energy scale, the superconducting gap, which is proportional
to the critical temperature ($T_c$)\cite{BCS}.

In hole-doped high-$T_c$ copper oxide superconductors,
previous experiments~\cite{Deutscher,LeTacon,Tanaka,Lee,Boyer}
have established the existence of two distinct energy scales for doping levels below the optimal one.
The origin and significance of these two scales are largely unexplained,
although they have often been viewed as evidence for two gaps, possibly
of distinct physical origins~\cite{Lee,Boyer,Millis,hufner}.
By measuring the temperature dependence of the electronic Raman response of
Bi$_{2}$Sr$_{2}$CaCu$_{2}$O$_{8+\delta}$ (Bi-2212) and HgBa$_{2}$CuO$_{4+\delta}$ (Hg-1201) crystals
with different doping levels,  we establish that
these two scales are associated with coherent excitations of the superconducting
state which disappears at $T_c$. Using a simple model, we show that these two scales
do not require the existence of two gaps. Rather, a single d-wave superconducting gap
with a loss of Bogoliubov quasiparticle spectral weight in the antinodal region is shown to reconcile spectroscopic
\cite{Guyard,Blanc,Devereaux,Kanigel1,Kondo,Chatterjee,McElroy,Yazdani,Kohsaka}
and transport \cite{HHWen,Matsuzaki,Sutherland,Hawthorn} measurements}.

In panels a-b of Fig.~1, we display the Raman spectra $\chi''(\omega,T)$ of Bi-$2212$ samples with different doping levels in $B_{1g}$  (antinodal) and $B_{2g}$ (nodal) geometries (see Methods), for several temperatures ranging from well below $T_c$ to $10$~K above $T_c$.
In both geometries, these spectra show the gradual emergence of
a peak as the sample is cooled below $T_c$.
An important observation is that the Raman responses
at $T_c$ and just above $T_c$ are essentially identical. This
suggests that the emergence of the \BAN and \BN peaks below $T_c$ is associated with
coherent excitations of the superconducting state.
The \BAN peak (measured at $T=10$~K)
is seen to decrease in intensity as the doping level is reduced,
and disappears altogether at the lowest doping $p\simeq 0.1$. In contrast, the
\BN peak intensity remains sizeable even at low doping levels.

In order to clearly reveal the temperature-dependence of the \BAN and \BN peaks,
we have plotted the normalized areas of the \BAN and \BN peaks in Fig.1e as a function of $T/T_c$.
This plot demonstrates that the peak intensities vanish at $T_c$ for all doping levels, providing quantitative support to our interpretation as coherence peaks of the superconducting state.
The subtracted spectra displayed in Fig.~1c-d (see caption) allow for a quantitative determination of
the characteristic energy scales associated with the superconducting
coherence peaks (see plot in Fig.~1f).
It is seen that the \BAN and \BN energy scales track each other at high doping levels ($p\gtrsim 0.19$), but
depart from each other as doping is reduced. The \BAN energy scale increases
monotonically as doping is reduced, while the \BN energy scale follows a dome-like
shape approximately similar to that of the critical temperature $T_c$.

In order to shed light on the origin of these two energy scales, we consider a
very simple phenomenological model of a superconductor with a gap function
$\Delta(\phi)$.
The angle $\phi$ is defined by
$\cos(2\phi)=\cos k_x - \cos k_y$ and the gap function vanishes at the nodal
point $\Delta(\phi=\pi/4)=0$ while it is maximal at the antinodes $\Delta(\phi=0)=\Deltam$.
Within a Fermi liquid description, the quasiparticle contribution to the
Raman response in the superconducting state reads~\cite{LeTacon,Devereaux}:
\begin{equation}\label{eq:response}
\chi^{''}_{\BAN,\BN}(\Omega)=
\frac{2\pi N_{F}}{\Omega}\left\langle \gamma^2_{\BAN,\BN}(\phi)\,Z^2(\phi)\Lambda^2(\phi)
\frac{\Delta(\phi)^2}{\sqrt{(\hbar\Omega)^2-4\Delta(\phi)^2}}\right\rangle_{FS}
\end{equation}
In this expression,
$\langle(\cdots)\rangle_{FS}$ denotes an angular average over the Fermi surface,
$\gamma_ {\BAN,\BN}$ are the Raman vertices which read
$\gamma_{\BAN}(\phi)=\gamma^{0}_{\BAN}\cos 2\phi$ and
$\gamma_{\BN}(\phi)=\gamma^{0}_{\BN}\sin 2\phi$, respectively.
, and $\Delta(\phi)^2/\sqrt{\Omega^2-4\Delta(\phi)^2}$ is a BCS coherence factor.
The function $Z(\phi)$ is the spectral weight of the Bogoliubov quasiparticles,
while $\Lambda(\phi)$ is a Fermi liquid parameter associated with the coupling of these
quasiparticles to the electromagnetic field.

In the following, we will show that the angular dependence of the quasiparticle renormalization, $Z.\Lambda(\phi)$,
plays a key role in accounting for the experimental observations.
In the \BAN geometry, the Raman vertex $\gamma_ {\BAN}(\phi)$
is peaked at the antinode $\phi=0$, resulting in
a pair-breaking coherence peak at $\hbar\Omega_{\BAN}=2\Deltam$ due to the singularity
of the BCS coherence factor. The weight of this peak is directly proportional to the
antinodal quasiparticle renormalization $(\ZAN)^2=(Z\Lambda)^2(\phi=0)$.
Hence, the fact that the \BAN coherence peak loses intensity at low doping strongly suggests that $\ZAN$ decreases
rapidly as doping is reduced, in qualitative agreement with tunneling \cite{McElroy,Kohsaka} and ARPES measurements \cite{Ding,Vishik}.

In the \BN geometry, the situation is more subtle because the Raman vertex is largest at
the nodes, where the BCS coherence factor vanishes.
As a result, the energy of the coherence peak depends sensitively on the angular
dependence of the quasiparticle renormalization $Z\Lambda(\phi)$. If the latter is
approximately constant along the Fermi surface, then the energy of the
\BN peak is determined solely by the angular extension of the Raman vertex $\gamma_{\BN}(\phi)$.
In contrast, let us consider a $Z\Lambda(\phi)$ which varies significantly, from a larger value \ZN
at the node to a small value \ZAN at the antinode, with a characteristic angular extension $\phi_N$ around the node,
smaller than the intrinsic width of the Raman vertex $\gamma_{\BN}(\phi)$. Then, it is $\phi_N$
itself which controls the position of the \BN peak: $\hbar\Omega_{\BN}=2\Delta(\phi_N)$.
As shown below, this key feature explains the origin of the differentiation between the two energy
scales in underdoped cuprates.

To proceed further in the simplest possible way, we consider a simple crenel-like shape  for
$Z\Lambda(\phi)$, varying rapidly from $\ZN$ for $\phi_N<\phi<\pi/4$ to
$\ZAN<\ZN$ for $0<\phi<\phi_N$ (Fig.~2.AII-CII).
Furthermore, we adopt the often-used~\cite{Mesot} parametrization
of the gap function $\Delta(\phi)=\Deltam\left[B\cos 2\phi + (1-B)\cos 6\phi\right]$,
consistent with d-wave symmetry where the nodal slope of the gap
$v_\Delta\equiv\partial\Delta/\partial\phi|_{\phi=\pi/4}=2(4B-3)\Deltam$ does not necessarily
track $\Deltam$.

We thus have 5 parameters: $\Deltam$, the nodal slope of the gap, \vD or $B$ (see Methods), \ZAN, \ZN and the angular extension \phiN.
These parameters are determined by attempting a semi-quantitative fit to our spectra, obeying
the following constraints:
\begin{itemize}
\item The maximum gap \Deltam is determined from the measured energy of the
\BAN peak according to $2\Deltam=\hbar\Omega_{\BAN}$
\item The antinodal quasiparticle renormalization \ZAN is determined such as to reproduce the
intensity of the \BAN coherence peak.
\item The angular extension \phiN is determined from the energy of
the nodal coherence peak. Throughout the underdoped regime, this amounts to
$2\Delta(\phi_N)=\hbar\Omega_{\BN}$ as discussed above.
\item The nodal renormalization \ZN is constrained to insure that the ratio
$(\ZN)^2/\vD$ does not change as a function of doping level, at least in the
range $0.1 < p < 0.16$ 
(as shown in Refs.~\cite{LeTacon,Blanc}). This ratio controls the low-frequency slope of the \BN Raman response.
We assume here that the density of states $N_F$ (associated with the Fermi velocity
perpendicular to the Fermi surface) does not depend sensitively on doping level in this range.
\end{itemize}

These 4 constraints leave one parameter undetermined, which can be taken as the deviation
of the gap function from a pure $\cos k_x-\cos k_y$ form, as measured by the ratio
$\vD/(2\Deltam)=4B-3$ of the nodal velocity to the gap maximum.
We will thus consider three possible scenarios:
\begin{itemize}
\item (A) Pure \cxcy gap: $\vD=2\Deltam$ ($B=1$). This corresponds to a superconducting gap
involving a single characteristic energy, which increases as the doping level is
reduced.
\item (B) \vD tracks the critical temperature $T_c$. In this case, the gap function is
truly characterized by two scales varying in opposite manner as the doping level is reduced.
\item (C) \vD remains constant as a function of doping. This is also a two-scale
superconducting gap scenario, although with a milder variation of \vD.
\end{itemize}

\par
In Fig.2, we display the \BAN and \BN Raman spectra calculated in the framework of
this simple theoretical analysis, following each of the three scenarios (A-C) above.
We observe that the main aspects of the experimental spectra, and most importantly
{\it the existence of two energy scales \OmAN, \OmN varying in opposite
manners as a function of doping}, can be reproduced within any of
the three scenarios.

A common feature between all three scenarios is that
the quasiparticle renormalization function
$Z\Lambda(\phi)$ varies significantly along the Fermi surface. Quasiparticles have a
large spectral weight \ZN only on a restricted region around the nodes, defined by $\phi_N$,
corresponding to a fraction $\arc\equiv (\pi/4-\phi_N)/(\pi/4)$ of the Fermi surface.
While $\Deltam$ increases with falling doping, $\Delta(\phi_N)$ decreases because of the rapid
contraction of the `coherent fraction' $\arc$, leading to the opposite doping dependence of the two scales, as illustrated on Fig.~3a.
We note that linearizing the gap function in the coherent region is a reasonable
approximation for the A and C scenarios, leading to the relation
$\hbar\OmN=\frac{\pi}{2}\arc\vD\propto k_BT_c$ which links the
nodal (\BN) energy scale (proportional to $T_c$), the nodal velocity and the coherent fraction.
This approximation is not valid for scenario B, because linearization is not accurate there.

It is clear that having uniformly coherent Bogoliubov quasiparticles along the Fermi surface is inconsistent with our data, especially in view of the rapid suppression of the
\BAN coherence peak and the corresponding decrease of \ZAN.
This is one of the key conclusions of the present work.

In panels (A.V-C.V) of Fig.2, we display a calculation of the corresponding tunneling conductance,
(see Methods).
One sees that the two energy scales have clear signatures in the tunneling spectra,
the nodal (\BN) one corresponding to a `kink'-like feature. Such a feature has indeed been observed in
tunneling spectra~\cite{Boyer,Yazdani,McElroy}, although it has been usually interpreted
as evidence that $\vD$ and $\Deltam$ are two distinct scales (scenarios B, or C),
without considering the effect of $Z(\phi)$.
Our work demonstrates that the STM `kink' results quite generally from the rapid decrease of the
quasiparticle spectral weight as one moves from nodes to antinodes.

\par
Although the above features are common to all three scenarios, there are
two key differences between them. The first one is qualitative:
in scenario A,
\ZN {\it increases} as doping level is reduced, while it {\it decreases} for scenario B
and stays constant for scenario C. The second, quantitative, difference is the rate at which the coherent fraction of the Fermi
surface \arc decreases with underdoping, being largest for scenario A and
smallest for B (Fig.~3b).

A determination of the coherent fraction $\arc^{HC}$ has been
reported from heat-capacity (HC) measurements~\cite{HHWen,Matsuzaki}, as reproduced
on Fig.~3b.
It was also reported from ARPES ~\cite{Kanigel1,Kanigel2} in the normal state, that the
Fermi arcs shrink upon cooling as $\sim T/T^*$. The doping evolution of the coherent fraction
$\arc^{ARPES}\propto T_c/T^*$ at $T_c$
is displayed in Fig.3b.
Remarkably, we find that there is a good {\it quantitative} agreement between the
doping dependence of \arc reported from HC and ARPES and our
determination from Raman within scenario A
(a single gap scale $\vD\propto\Deltam$), which thus appears to be favored by this comparison.
Although this quantitative agreement should perhaps not be overemphasized in view
of the uncertainties associated with each of the experimental probes, we conclude that
this single-gap scenario (A) stands out as the most likely possibility.
Our interpretation also reconciles the distinct doping
dependence of the two energy scales with thermal conductivity measurements in
underdoped samples which, interpreted within the clean limit, suggest that
$\vD\propto\Deltam$ \cite{Sutherland,Hawthorn} (see, however \cite{Sun}).

\par
As noted before, with a single superconducting gap, the relation between the critical temperature
(or $\OmN$) and the coherent fraction reads: $k_BT_c \propto \arc\Deltam$.
We are then led to conclude that it is the suppressed
coherence of the quasiparticles that sets the value of $T_c$, while $\Deltam$
increases with underdoping.
From our findings, we anticipate that the loss of coherent of Bogoliubov quasiparticles on a restricted part of the Fermi surface only is a general feature of how superconductivity emerges as holes are doped into a Mott insulating state.

\section*{METHODS}
\subsection*{Details of the experimental procedure}

The Bi-2212 and Hg-1201 single crystals have been grown by using a floating zone method and flux technique respectively. The detailed procedures of the
crystal growth are described elsewhere~\cite{Gu,Blanc,Bertinotti}. The doping value $p$ is inferred from $T_c$ using equation of Presland et al from Ref. ~\cite{PreslandPhysicaC91}: $1-T_c/T_{c}^{max} = 82.6\,(p-0.16)^2$. $T_c$ has been determined from magnetic susceptibility measurements for each doping level.
Raman experiments have been carried out using a triple grating spectrometer (JY-T64000) equipped with a nitrogen cooled CCD detector.
All the measurements have been corrected for the Bose factor and the instrumental spectral response. The $B_{1g}$ and $B_{2g}$ geometries have been obtained using crossed light polarizations at 45$^o$ from the $Cu-O$ bond directions and along them respectively.
In these geometries we probe respectively, the antinodal (AN) and nodal (N) regions correponding to the principal axes and the diagonal of the Brillouin zone.

\subsection*{Calculation of the tunneling conductance}

$\frac{dI}{dV} \propto N_F \left\langle\, Z(\phi)\,
\rm{Re} \frac{eV-i\Gamma}{(eV-i\Gamma)^2-\Delta(\phi)^2}\right\rangle_{FS}$
where $\Gamma=\Deltam/30$ is a small damping factor, and we have assumed for simplicity
that $Z(\phi)$ and $Z\Lambda(\phi)$ can be identified.

\section*{ACKNOWLEDGEMENTS}
We are grateful to  J. Mesot, A. Yazdani, A.~J.~Millis, P.~Monod and C. Ciuti for very helpful discussions.
Correspondance and requests for materials should be addressed to A.S. (alain.sacuto@univ-paris-diderot.fr).



\newpage
\begin{figure}
\begin{center}
\includegraphics[width=18cm]{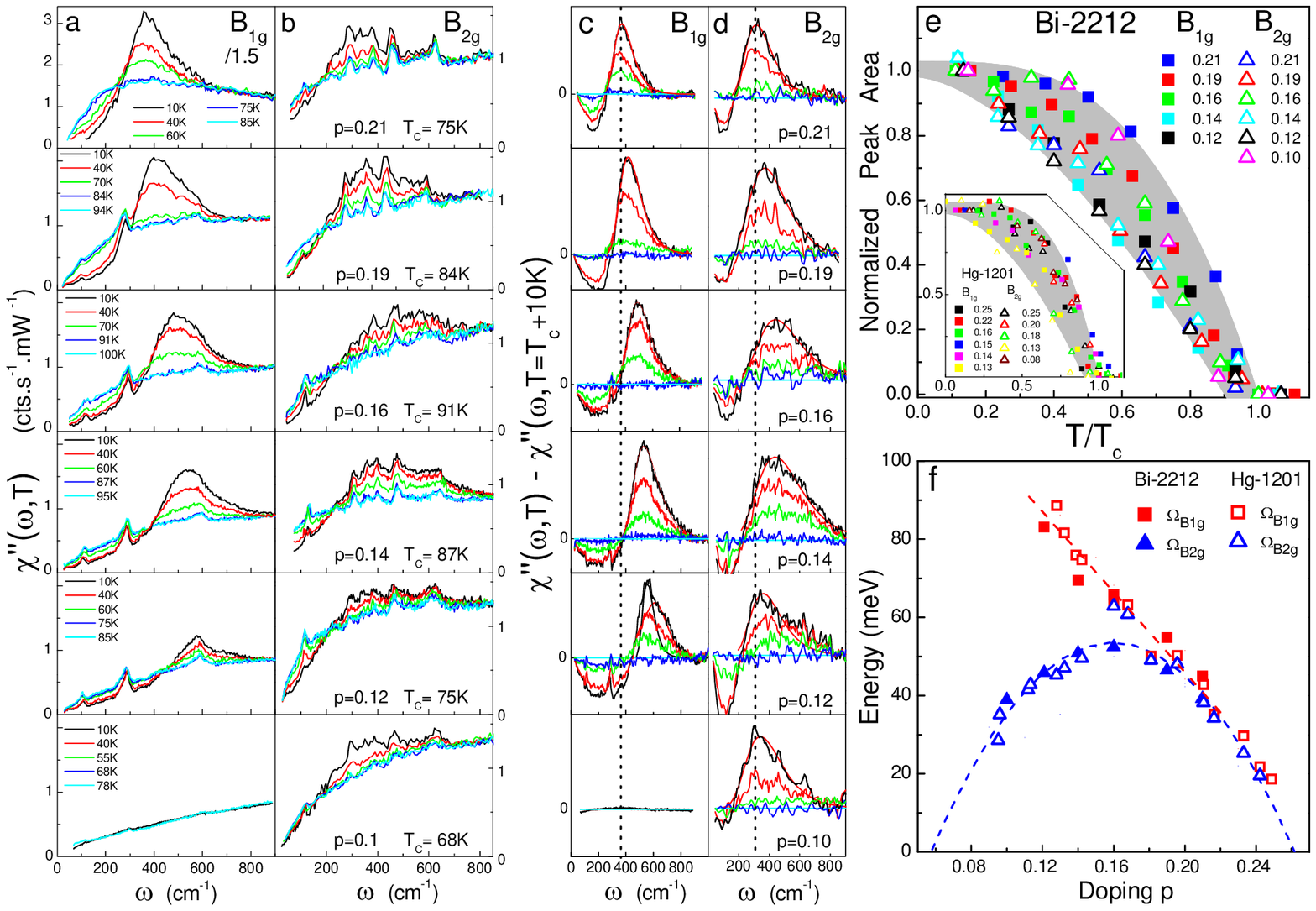}

\end{center}\vspace{-7mm}
\caption{(color online) \textbf{Temperature dependence of the Raman spectra.}
\textbf{a},\textbf{b}: Raman spectra, $\chi''(\omega,T)$ of  Bi-$2212$ single crystals for several doping levels in $B_{1g}$ (Antinodal) and $B_{2g}$ (Nodal) geometries.
\textbf{c}, \textbf{d}: Raman spectra subtracted from the one measured at $10~K$ above $T_c$ for
each sample in each geometry.  A direct visual comparison of the subtracted spectra to a reference
energy (chosen as the peak position for the most overdoped sample, drawn as a guide to the eyes) clearly
reveals the distinct doping-dependence of these two energy scales. \textbf{e}: Temperature dependence of the normalized areas of the $B_{1g}$ and $B_{2g}$ peaks with respect to the area measured at $T=10~K$. The inset displays these areas for Hg-$1201$ crystals.
\textbf{f}: Energy scales associated to the $B_{1g}$ and $B_{2g}$ peaks for both Bi-2212 and Hg-1201 crystals.
The dashed lines are guides to the eyes and track the locations of the superconducting peak maxima.}
\label{fig1}
\end{figure}

\newpage
\begin{figure}
\begin{center}
\includegraphics[width=12cm]{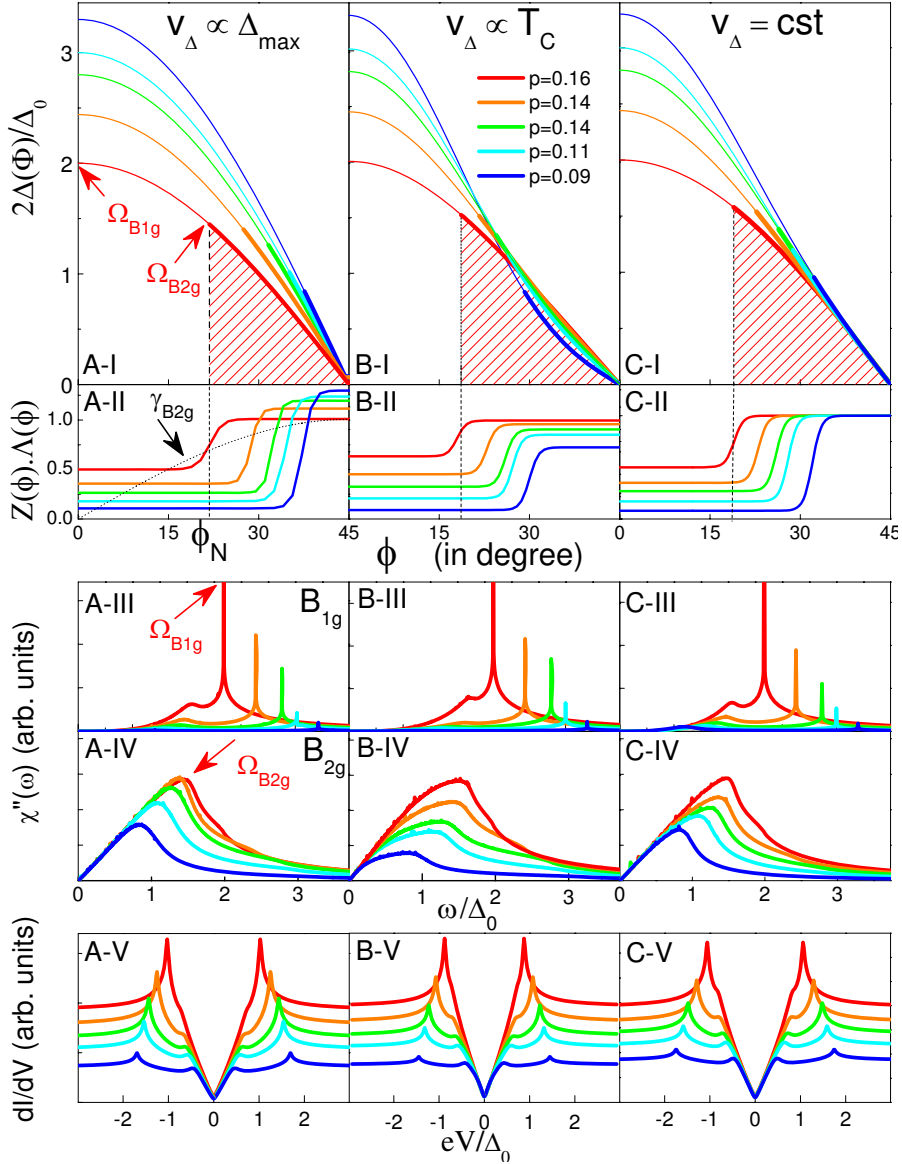}
\end{center}\vspace{-7mm}
\caption{(color online) \textbf{Three scenarios for underdoped cuprates.}
\textbf{A.I-C.I}: Doping evolution of the superconducting gap in the three scenarios (A-C).
\textbf{A.II-C.II}: Angular dependences of the quasiparticle spectral weights $Z\Lambda(\phi)$
as a function of doping level for each scenario. The angular dependence of the $B_{2g}$ Raman vertex
is shown in dotted line (see \textbf{AII}).
\textbf{A.III-C.III, A.IV-C.IV}: Calculated Raman spectra for each scenario in $B_{1g}$ and $B_{2g}$ geometries. The locations of the $B_{1g}$ and $B_{2g}$ peaks are respectively controlled by the gap energy at the antinodes $\Phi=0$ and at the angle $\Phi_{N}$.
Note that the low energy slope of the $B_{2g}$ Raman response is constant with doping as observed
experimentally [\onlinecite{Blanc}].
\textbf{A.V-C.V}: Tunneling spectra for the three scenarios.}
\label{fig4}
\end{figure}

\newpage
\begin{figure}
\begin{center}
\includegraphics[width=18cm]{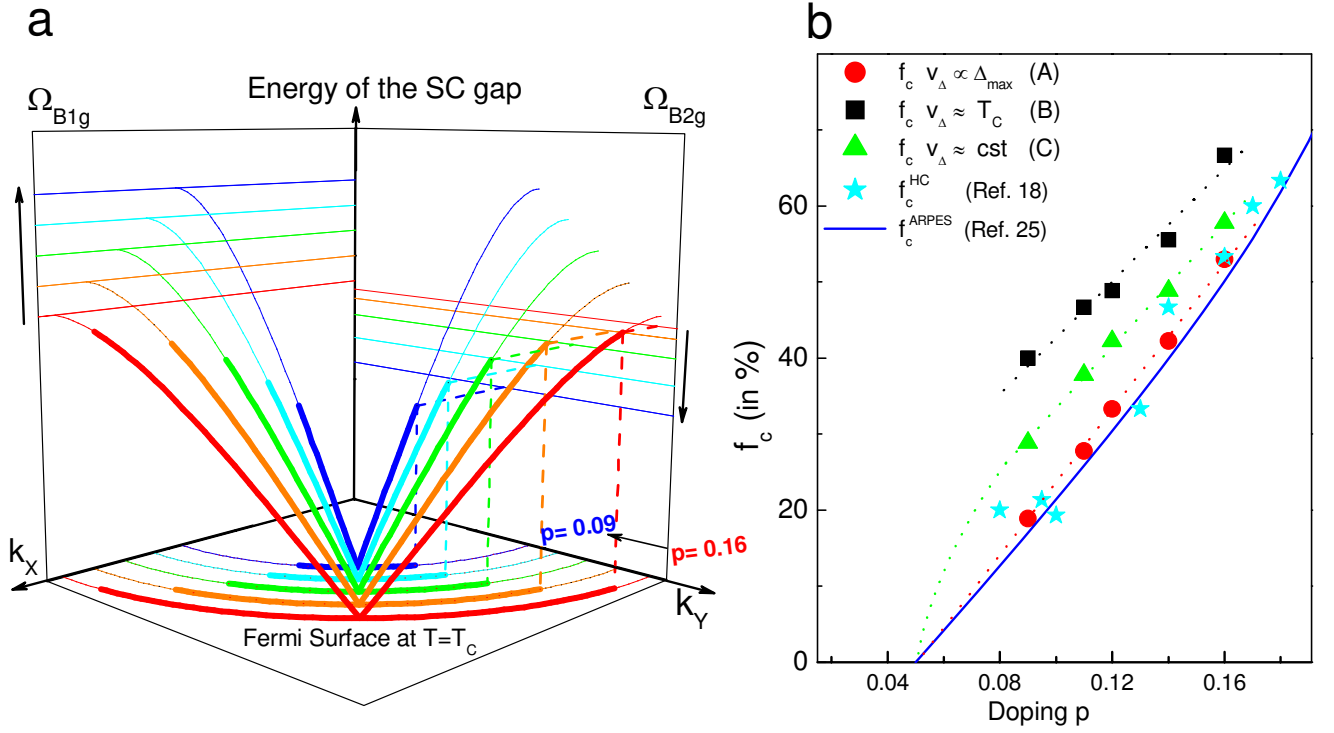}
\end{center}\vspace{-7mm}
\caption{(color online) \textbf{Fermi surface coherent fraction and a single d-wave superconducting gap.} \textbf{a}: Scenario A where coherent Bogoliubov quasiparticles are partially suppressed on restricted parts of the Fermi surface and the superconducting gap has a single d-wave shape. \textbf{b}: Doping evolution of the coherent fraction of the Fermi surface ($\arc$) in the three scenarios (A-C). The $\arc^{ARPES}$ curve is deduced from $L_{arc}/L_{full}(T_{c})=1-0.70\frac{T_{c}}{T^{*}}$
(see Ref.[\onlinecite{Kanigel2}]), expressed as a function of the doping level. $\arc^{HC}$ is extracted from the zero temperature specific heat coefficient $\gamma(0)_{n}$,  [\onlinecite{HHWen,Matsuzaki}].}

\label{fig3}
\end{figure}

\end{document}